\documentclass[aps,pra,showpacs,twocolumn,superscriptaddress]{revtex4-1}
\usepackage{graphicx,amsmath,amssymb}
\usepackage[usenames]{color}
\usepackage{indentfirst}
\usepackage{float}

\usepackage[colorlinks=true, citecolor=blue, urlcolor=blue, linkcolor=blue ]{hyperref}
\hypersetup{breaklinks=true}
\begin{document}
\bibliographystyle{apsrev4-1}
\title{Microwave-controlled optical double optomechanically induced transparency in a hybrid piezo-optomechanical cavity system}
\author{Shi-Chao Wu}
\affiliation{Shanghai Advanced Research Institute, Chinese Academy of Sciences, Shanghai, 201210, China}
\affiliation{University of Chinese Academy of Sciences, Beijing, 100049 ,China}
\author{Li-Guo Qin}
\affiliation{Shanghai Advanced Research Institute, Chinese Academy of Sciences, Shanghai, 201210, China}
\affiliation{School of Science, Qingdao University of Technology, Qingdao, 266000, Shandong, China}
\author{Jun Jing}
\affiliation{Department of Physics, Zhejiang University, Hangzhou, 310027, Zhejiang, China}
\author{Tian-Min Yan}
\affiliation{Shanghai Advanced Research Institute, Chinese Academy of Sciences, Shanghai, 201210, China}
\author{Jian Lu}
\affiliation{Shanghai Advanced Research Institute, Chinese Academy of Sciences, Shanghai, 201210, China}
\author{Zhong-Yang Wang}
\email{wangzy@sari.ac.cn}
\affiliation{Shanghai Advanced Research Institute, Chinese Academy of Sciences, Shanghai, 201210, China}

\begin{abstract}
  We propose a scheme that can generate a microwave-controlled optical double optomechanical induced transparency (OMIT) in a hybrid piezo-optomechanical cavity system, in which a piezoelectric optomechanical crystal AlN-nanobeam resonator is placed in a superconducting coplaner microwave cavity. We show that when an intense microwave field is applied to the superconducting microwave cavity, with a strong pump optical field and a weak probe optical field applied to the optomechanical crystal cavity through the optical waveguide, a double transmission window can be obtained in the weak output probe field. This phenomenon arises because an N-type four-level system can be formed in our scheme, under the effects of the radiation pressure and the piezoelectric interaction, the quantum interference between different energy level pathways induces the occurrence of the double-OMIT. Our scheme can be applied in the fields of optical switches, high-resolution spectroscopy, coherent population trapping and quantum information processing in solid-state quantum systems.
\end{abstract}


\maketitle

\section{\label{sec:level1} INTRODUCTION}
Mutual interaction between the optical field and the microwave field is an active research area, that has been studied theoretically and experimentally in many fields, such as the coherent signal transfer \cite{PRLStateTransferwang,PRLinterfaceBarzanjeh,PRAtransferMcGee}, bidirectional conversion \cite{andrewsNaturebidirectionalconver,Natureultrafastconversion,PRAconversionXu} and coherent coupling \cite{naturenanomechanicalBochmann,PRAEntanglingBarzanjeh,NatCoherentBalram,NatureQuantumcoherent}. Those phenomena have been realized in many systems, a representative example of which is the optomechanical system. A traditional optomechanical system comprises mechanical resonators and optical or microwave field cavities \cite{RevMooptomechanicsAspelmeyer}. With the advantage of integration, the traditional optomechanical system has been extended to various hybrid solid-state quantum systems, including integration with optical lattice crystals \cite{NatCoherentBalram}, piezoelectric materials \cite{PRAppliedAcoustoOpticBalram}, superconducting microwave cavities \cite{regalnaturemeasuring}, and superconducting quantum circuits \cite{RMPHybridquantumXiang}.

Since the electromagnetically induced transparency (EIT) phenomenon was observed in three-level atomic systems \cite{PhysicsTodayHarris,PRLEITBoller}, many novel EIT-related effects have also been studied theoretically and experimentally in multiple-level systems, such as the two-photon absorption phenomenon \cite{PRLNonlinearAndr,PRLPhotonHarris,OLNONABSORPMin,PRLNONLINEKang,PRAEITJiang,PRAMagnetoPetrosyan,sedlacekNaturemicrowaveNlevel}. Two-photon absorption was first proposed theoretically by Harris \emph{et al} \cite{PRLPhotonHarris} and observed by Yan \emph{et al} \cite{OLNONABSORPMin} in the Rb atomic system; this effect arises from quantum interference occurring in the four-level energy structure formed by the atomic systems. One important application of two-photon absorption is to realize the double-EIT phenomenon, which can be used in the fields of high-speed optical switches \cite{PRAswitchingKumar}, high-resolution spectroscopy \cite{PRLNonlinearHarris}, cross-phase modulation \cite{PRLCrossPhaseLi}, and quantum information processing \cite{PRLPairedphotonsBali}.

In addition to EIT in natural atomic systems, optomechanical induced transparency (OMIT), which is an EIT-like phenomenon, has also been studied in various optomechanical systems \cite{NatCoherentBalram,RMPHybridquantumXiang,ScienceOptomechanicallyWeis,safavinatureelectromagnetically,PRATunableMa,
PRAOptomechanicalWang,PRAPhaseJia,PRAPrecisionZhang,PRAMicrowaveassistedKingHan,PRAcoupledresonatorsDuan,PRBHybridquantumWangx}, such as the optomechanical crystal nanobeam cavity system \cite{safavinatureelectromagnetically}. An optomechanical crystal nanobeam cavity system is formed based on a freestanding beam structure, in which an array of periodic elliptical air holes is patterned. The system is designed to support the co-localized high-Q optical cavity mode and the mechanical resonance mode simultaneously. In this device, the optical mode and the mechanical mode can interact strongly with each other via radiation pressure; a single-photon radiation pressure coupling strength of $1.1\times10^6$ Hz has been realized experimentally \cite{NatCoherentBalram}.

Recently, in some research experiments, optomechanical crystal nanobeams have been fabricated using both photoelastic and piezoelectric materials \cite{NatCoherentBalram,PRAppliedAcoustoOpticBalram,naturenanomechanicalBochmann}, such as AlN \cite{naturenanomechanicalBochmann}. Under the effects of the radiation pressure and the piezoelectric interaction, the AlN-nanobeam resonator can be driven by both the optical field and the microwave field simultaneously. Strong piezoelectric coupling between the microwave field and the mechanical resonator has also been realized recently \cite{PRACavityChangLZ,PRLMultimodeHan,PRLSuperconductingCleland,naturequantumConnell}, for example, in the superconducting coplanar microwave cavity system, the piezoelectric coupling strength between the microwave and mechanical modes can be designed to reach $12.3\times10^6$ Hz \cite{PRACavityChangLZ}.

Here, we propose a hybrid optical and microwave piezo-optomechanical cavity system, in which an optomechanical crystal AlN-nanobeam resonator is placed in a superconducting coplanar microwave cavity without contact. In this system, the AlN-nanobeam mechanical resonator can be effectively driven by both the microwave field and the optical field simultaneously, under the effects of the radiation pressure and the piezoelectric interaction, respectively. We show that, when an intense microwave field is applied to the superconducting microwave cavity, with a strong pump and a weak probe optical field applied to the optomechanical crystal cavity through the optical waveguide, a double-OMIT window can be obtained in the weak output probe field. Similar to the double-EIT observed in the natural atomic systems, the double-OMIT can be applied in the fields of optical switches, high-resolution spectroscopy, and quantum information processing in solid-state quantum systems.

This phenomenon arises because an N-type four-level structure can be formed in our scheme, similar to the double-EIT phenomenon, which also arises from the four-level structure formed in the natural atomic systems, and the relevant mechanisms have been studied extensively \cite{PRAObservationYan,PRLCrossPhaseLi}. In contrast to the single-OMIT phenomenon observed in the traditional $\Lambda$-type three-level optomechanical system \cite{PRAEITAgarwal}, the energy level of the mechanical resonator is split into two new dressed levels in our system. Under the effect of quantum interference between different energy level pathways, the third-order nonlinear absorption can be enhanced by constructive quantum pathway interference while the linear absorption is inhibited by destructive quantum pathway interferencel; as a result, the single transmission window is split into two new windows to generate the double-OMIT window.

The double-OMIT phenomenon has been studied theoretically in many three-mode nonlinear optomechanical systems \cite{PRA2013nonlinearShahidani,JPB2014DoubleSumei,PRA2013PhononmediatedQu,PRATunableMa,PRAOptomechanicalWang,OC2014DoubleWen}, according to the different types of combinations, these systems can be divided into the following classes: the cavity mode coupled with two mechanical resonator modes \cite{PRA2013nonlinearShahidani,JPB2014DoubleSumei}, the mechanical resonator mode coupled with two cavity modes \cite{PRA2013PhononmediatedQu}, the mechanical resonator mode coupled with a cavity mode and an another mechanical resonator \cite{PRATunableMa} (or a qubit \cite{PRAOptomechanicalWang}) mode, the cavity mode coupled with a mechanical resonator mode and another cavity mode \cite{OC2014DoubleWen}, and so on. Relative to the previously studied systems, our system has the following advantages: (i) our proposed scheme is a combination of the piezoelectric optomechanical crystal AlN-nanobeam resonator and the superconducting coplanar microwave cavity, which is relatively robust and scalable; (ii) the distance between the two windows of the double-OMIT can be controlled by changing the piezomechanical coupling strength, which is related to the position of the AlN-nanobeam resonator in the microwave cavity; (iii) our scheme can realize switching between the double-OMIT and the single-mode OMIT by changing the frequency of the microwave field, and the transmission width of the single-mode OMIT window can be controlled by changing the power of the microwave field; (iv) moreover, our scheme can also realize the switching between the symmetric and asymmetric double-OMIT window by changing the frequency of the microwave field, in contrast to previous schemes, in which the frequencies of the mechanical resonator \cite{PRATunableMa} and qubit \cite{PRAOptomechanicalWang} are fixed in the designed devices, only the symmetric or asymmetrical double-OMIT window is generated.

The paper is organized as follows. In Sec. II we describe the proposed system and derive the quantum Langevin equations(QLEs). In Sec. III we discuss the experimental feasibility of our system and the physical mechanism of the double-OMIT. In Sec. IV the tunable-OMIT controlled by the inputs fields is presented. The last section concludes the paper.

\section{\label{sec:level1} THEORETICAL MODEL}
The schematic of our proposed hybrid optical and microwave piezo-optomechanical cavity system is shown in Fig.1(a), in which an optomechanical crystal AlN-nanobeam resonator is placed in a superconducting coplanar microwave cavity without contact. Fig.1(b) shows the schematic diagram of the experimental configuration. The optomechanical crystal cavity AlN-nanobeam resonator is a mechanically suspended AlN-beam, in which an array of periodic elliptical air holes is patterned. The resonator is designed to support the co-localized high-Q optical cavity modes and the mechanical resonance modes simultaneously, and the same structure was fabricated and applied in a related experiment \cite{naturenanomechanicalBochmann}. The superconducting coplanar microwave cavity comprises the top electrode, bottom electrodes, SiO$_2$ substrate and silicon substrates, where the bottom electrode is offset from the AlN-nanobeam resonator to form the ground plane, and the top electrode is fabricated beside the AlN-nanobeam resonator with a gap between them. This arrangement can ensure a considerable portion of the out-of-plane electric field in the AlN-nanobeam resonator and ease fabrication challenges \cite{PRACavityChangLZ}. The AlN-nanobeam mechanical resonator can be effectively driven by both the microwave field and the optical field, in which the microwave field is applied to the superconducting coplanar microwave cavity directly, and the optical fields are applied to the system via the optical waveguide. The coupling between the optical field and the mechanical motion is referred to as the optomechanical coupling, and the coupling between the microwave field and the mechanical motion is referred to as the piezomechanical coupling. More importantly, as a gap exists between the top electrode and the AlN-nanobeam resonator, the piezoelectric coupling strength can be adjusted by changing the distance between them, with the strength being inversely proportional to the distance \cite{PRACavityChangLZ}.

\begin{figure}[tbp]
\begin{center}
\includegraphics[width=0.9\columnwidth]{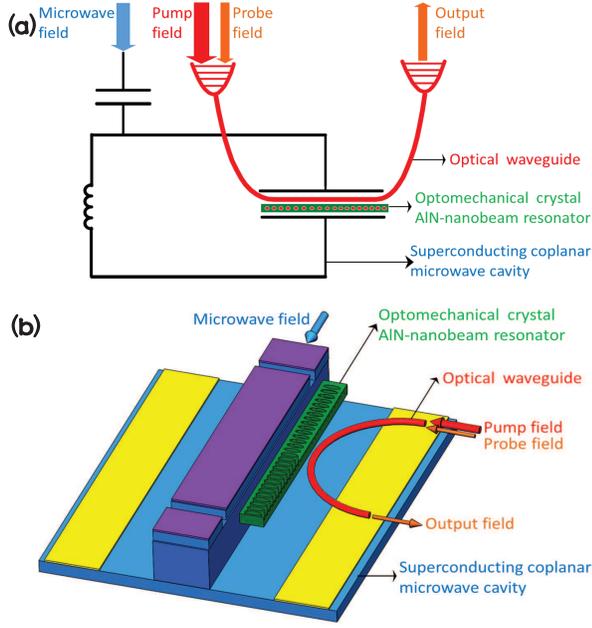}
  \caption{(color online) (a) Schematic diagram of the hybrid optical and microwave piezo-optomechanical cavity system, in which an optomechanical crystal AlN-nanobeam resonator is placed in a superconducting coplanar microwave cavity without contact. The superconducting coplanar microwave cavity is driven by an intense microwave field. The optomechanical crystal AlN-nanobeam resonator is driven by a strong pump field and a weak probe field via the optical waveguide. (b) Schematic diagram for the experimental configuration, where the superconducting coplanar microwave cavity comprises a top electrode (purple), bottom electrodes (yellow), SiO$_2$ substrate (cyan) and silicon substrates (blue).}
\end{center}
\end{figure}

We assume that when the superconducting planar microwave cavity is driven by an intense microwave field with frequency $\omega_k$, a strong optical pump field with frequency $\omega_{pu}$ and a weak optical probe field with frequency $\omega_{pr}$ are applied to the optomechanical AlN-nanobeam resonator cavity via the optical waveguide. The frequency of the microwave cavity is $\omega_c$, and the frequencies of the co-localized optical cavity mode and mechanical phonon mode are $\omega_a$ and $\omega_b$, respectively. By adopting the interaction picture with respect to $H^{\prime}=\hbar \omega_{pu}{\hat a}^{\dag} \hat{a}+\hbar \omega_k{\hat c}^{\dag} \hat{c}$, the Hamiltonian of the total system can be given by
\begin{eqnarray}
H=H_{0}+H_{I},
\end{eqnarray}
where
\begin{eqnarray}
H_0=\hbar{\Delta}_{{a}}{\hat{a}}^{\dag}a+\hbar{\Delta}_{{{c}}}{{\hat{c}}}^{\dag}\hat{c}+\omega_{b}{\hat{b}}^{\dag}\hat{b},
\end{eqnarray}
\begin{eqnarray}
H_I=-\hbar{g_{om}}\hat{a}^{\dag}\hat{a}(\hat{b}^{\dag}+\hat{b})-\hbar{g_{em}}(\hat{b}^{\dag}\hat{c}+\hat{b}\hat{c}^{\dag}) \nonumber \\
+({i}{\hbar}{\varepsilon}_{pu}\hat{a}^{\dag}+{i}{\hbar}{\varepsilon}_ {pr}e^{{-i}{\delta}t}{\hat{a}}^{\dag}+{i}{\hbar}{\varepsilon}_{k}\hat{c}^{\dag}+H.c.).
\end{eqnarray}
$H_0$ is the free Hamiltonian of the system, where $\hat{a}$, $\hat{c}$ and $\hat{b}$ are the annihilation operators of the optical cavity mode, the microwave cavity mode and the mechanical phonon mode, respectively. $\Delta_a=\omega_a-\omega_{pu}$ is the detuning of the optical pump field from the optical cavity, and $\Delta_c=\omega_c-\omega_{k}$ is the detuning of microwave driven field from the microwave cavity. $H_I$ describes the interaction between the hybrid piezo-optomechanical cavity system and the input fields. The first term is the optomechanical interaction term. $g_{om}$ denotes the single-photon optomechanical coupling strength between the microwave field and the mechanical phonon mode and is defined as $g_{om}=({\omega_a}/{L})\sqrt{\hbar/m\omega_b}$, where $m$ is the effective mass of mechanical mode and $L$ is the effective length of the optical cavity. The second term is the piezomechanical interaction term, where $g_{em}$ is the piezomechanical coupling strength between the microwave field and the mechanical phonon mode. The last four terms describe the energy of the input fields, where $\delta=\omega_{pr}-\omega_{pu}$ is the detuning of the optical probe field from the pump field. The intensities of the input optical pump field, optical probe field, and the microwave field are defined as $\varepsilon_{pu}={\sqrt{P_{pu}\kappa_{a}/(\hbar\omega_{pu})}}$, $\varepsilon_{pr}={\sqrt{P_{pr}\kappa_{a}/(\hbar\omega_{pr})}}$ ,and $\varepsilon_{k}={\sqrt{P_{k}\kappa_{c}/(\hbar\omega_{k})}}$, respectively, where $P_{pu}$, $P_{pr}$, and $P_{k}$ are the input powers of the optical pump field, optical probe field, and the microwave field, respectively. $\kappa_a$ and $\kappa_c$ are the decay rates of the optical cavity and the microwave cavity, respectively.

However, the dynamics of the three modes are also affected by damping and noise processes. By adopting the approach involving the QLEs, in which the Heisenberg equations for the operators are supplemented with damping and noise terms \cite{PRAPhaseJia,PRAEntanglingBarzanjeh}, we find the resulting nonlinear QLEs to be
\begin{eqnarray}
\begin{split}
\dot{\hat{a}}=-(i\Delta_{a}+\frac{\kappa_a}{2})\hat{a}+ig_{om}{\hat{a}}({\hat{b}}^{\dag}+{\hat{b}})
\\+\varepsilon_{pu}+\varepsilon_{pr}{e^{-i{\delta t}}}+\hat{f}_{in},\\
\dot{\hat{c}}=-(i\Delta_{c}+\frac{\kappa_c}{2}){\hat{c}}+ig_{em}{\hat{b}}+\varepsilon_{k}+\hat{\xi}_{in},\\
\dot{{\hat{b}}}=-(i\omega_{b}+\frac{\gamma_{b}}{2}){\hat{b}}+ig_{om}{\hat{a}^\dag}\hat{a}+ig_{em}{\hat{c}}+\hat{\varsigma}_{in},
\end{split}
\end{eqnarray}
where $\gamma_{b}$ is the intrinsic damping rate of mechanical resonator; $\hat{f}_{in}$ and $\hat{\xi}_{in}$ are the optical and microwave input noises, respectively; and $\hat{\varsigma}_{in}$ is the quantum Brownian noise acting on the nanobeam resonator \cite{PRAEntanglingBarzanjeh}. For simplicity, the hat symbols of the operators are omitted in the rest of this description.

Relative to the intensities of the optical pump field and the microwave field, the optical probe field is a weak field that satisfies the conditions $\varepsilon_{pr}\ll\varepsilon_{pu}$ and $\varepsilon_{pr}\ll\varepsilon_{k}$. We can linearize the dynamical equations of the system by assuming $a={a}_{s}+\delta{a}$, $c={c}_{s}+\delta{c}$ and $b={b}_{s}+\delta{b}$, all of which are composed of an average amplitude and a fluctuation term. Here, ${a}_{s}$, ${c}_{s}$ and ${b}_{s}$ are steady-state values when only the strong optical pump field and the microwave field are applied. By assuming $\varepsilon_{pr}\rightarrow0$ and setting all the time derivatives to zero, we obtain
\begin{eqnarray}
\begin{split}
{a}_{s}=\frac{\varepsilon_{pu}}{i\Delta_{a}^{\prime}+\frac{\kappa_a}{2}},\\
{c}_{s}=\frac{ig_{em}{b}_{s}+\varepsilon_{k}}{i\Delta_{c}+\frac{\kappa_c}{2}},\\
{b}_{s}=\frac{ig_{om}{|a_s|}^2+ig_{em}{c}_{s}}{i\omega_{b}+\frac{\gamma_{b}}{2}},
\end{split}
\end{eqnarray}
where $\Delta_{a}^{\prime}=\Delta_{a}-g_{om}( {b}_{s}^{\ast}+{b}_{s} )$, $\Delta_{a}^{\prime}$ is the effective detuning of the optical pump field from the optical cavity, including the frequency shift caused by the mechanical motion. Furthermore, by substituting the assumptions $a={a}_{s}+\delta{a}$, $c={c}_{s}+\delta{c}$ and $b={b}_{s}+\delta{b}$ into the nonlinear QLEs and dropping the small nonlinear terms, we can obtain the linearized QLEs as follows:
\begin{eqnarray}
\begin{split}
\dot{\delta{{a}}}=-(i\Delta_a^{\prime}+\frac{\kappa_a}{2})\delta{a}+iG_{om}(\delta{b^{\dag}}+\delta{b})+\varepsilon_{pr}e^{-i\delta{t}}+f_{in},\\
\dot{\delta{{c}}}=-(i\Delta_{c}+\frac{\kappa_c}{2})\delta{c}+ig_{em}\delta{b}+\xi_{in},\\
\dot{\delta{{b}}}=-(i\omega_{b}+\frac{\gamma_{b}}{2})\delta{b}+i(G_{om}^{\ast}{\delta{a}}+G_{om}\delta{a^{\dag}})+ig_{em}\delta{c}+\varsigma_{in},
\end{split}
\end{eqnarray}
where $G_{om}=g_{om}{{a_{s}}}$ is the total coupling strength between the optical mode and mechanical mode.

We assume that the system is operated in the resolved sideband regime, in which $\omega_b\gg\kappa_a$ and $\omega_{b}\gg{\kappa_{c}}$. The mechanical resonator has a high quality factor for $\omega_{b}\gg\gamma_b$, and the mechanical frequency $\omega_{b}$ is also much larger than $G_{om}$ and $g_{em}$. The fluctuation terms $\delta{a}$, $\delta{c}$, $\delta{b}$ and the noise terms ${f_{in}}$, ${\xi_{in}}$, ${\varsigma_{in}}$ can be rewritten as
\begin{eqnarray}
\begin{split}
\delta{a}=\delta{a_{+}}e^{-i{\delta}t}+\delta{a_{-}}e^{i{\delta}t},\\
\delta{c}=\delta{c_{+}}e^{-i{\delta}t}+\delta{c_{-}}e^{i{\delta}t},\\
\delta{b}=\delta{b_{+}}e^{-i{\delta}t}+\delta{b_{-}}e^{i{\delta}t},\\
\delta{f_{in}}=\delta{{f_{in}}_{+}}e^{-i{\delta}t}+\delta{{f_{in}}_{-}}e^{i{\delta}t},\\
\delta{\xi_{in}}=\delta{{\xi_{in}}_{+}}e^{-i{\delta}t}+\delta{{\xi_{in}}_{-}}e^{i{\delta}t},\\
\delta{\varsigma_{in}}=\delta{{\varsigma_{in}}_{+}}e^{-i{\delta}t}+\delta{{\varsigma_{in}}_{-}}e^{i{\delta}t},
\end{split}
\end{eqnarray}
where $\delta{O_{+}}$ and $\delta{O_{-}}$ (with $O=a,c,b$) correspond to the components at the original frequencies of $\omega_{pr}$ and $2\omega_{pu}-\omega_{pr}$, respectively \cite{PRAEITSumei,PRAPrecisionZhang}. Substituting Eq. (7) into Eq. (6) and ignoring the second-order small terms and equating coefficients of terms with the same frequency, we obtain the component at the original frequencies $\omega_{pr}$ as follows:
\begin{eqnarray}
\begin{split}
\dot{\delta{{a_{+}}}}=(i\lambda_a-\frac{\kappa_a}{2})\delta{a_{+}}+iG_{om}\delta{b_{+}}+\varepsilon_{pr}+{f_{in}}_{+},\\
\dot{\delta{{c_{+}}}}=(i\lambda_c-\frac{\kappa_c}{2})\delta{c_{+}}+ig_{em}\delta{b}_{+}+{\xi_{in}}_{+},\\
\dot{\delta{{b_{+}}}}=(i\lambda_{b}-\frac{\gamma_{b}}{2})\delta{b_{+}}+iG_{om}^{\ast}{\delta{a}}_{+}+ig_{em}\delta{c}_{+}+{\varsigma_{in}}_{+},
\end{split}
\end{eqnarray}
where $\lambda_a=\delta-\Delta_{a}^{\prime}$, $\lambda_c=\delta-\Delta_{c}$, and $\lambda_{b}=\delta-\omega_{b}$.

Next, we take the expectation values of the operators in Eq. (8). As the noise terms obey the following correlation fluctuations\cite{RevMooptomechanicsAspelmeyer}:
\begin{eqnarray}
\begin{split}
\langle{{{{\hat{f}}_{in}(t) \hat{f}}_{in}^{\dag}(t^{\prime})}}\rangle=[N(\omega_{a})+1]\delta(t-t^{\prime}),\\
\langle{{{{\hat{f}}_{in}^{\dag}(t) \hat{f}}_{in}(t^{\prime})}}\rangle=[N(\omega_{a})]\delta(t-t^{\prime}),\\
\langle{{{{\hat{\xi}}_{in}(t) \hat{\xi}}_{in}^{\dag}(t^{\prime})}}\rangle=[N(\omega_{c})+1]\delta(t-t^{\prime}),\\
\langle{{{{\hat{\xi}}_{in}^{\dag}(t) \hat{\xi}}_{in}(t^{\prime})}}\rangle=[N(\omega_{c})]\delta(t-t^{\prime}),\\
\langle{{{{\hat{\varsigma}}_{in}(t)\hat{\varsigma}}_{in}^{\dag}(t^{\prime})}}\rangle=[N(\omega_{b})+1]\delta(t-t^{\prime}),\\
\langle{{{{\hat{\varsigma}}_{in}^{\dag}(t)\hat{\varsigma}}_{in}(t^{\prime})}}\rangle=[N(\omega_{b})]\delta(t-t^{\prime}),
\end{split}
\end{eqnarray}
where $N(\omega_{a})=[exp(\hbar\omega_{a}/k_{B}T)-1]^{-1}$ and $N(\omega_{c})=[exp(\hbar\omega_{c}/k_{B}T)-1]^{-1}$
are the equilibrium mean thermal photon numbers of the optical and microwave fields, respectively; $N(\omega_{b})=[exp(\hbar\omega_{b}/k_{B}T)-1]^{-1}$ is the equilibrium mean thermal phonon number of the nanobeam resonator. We can safely assume that the optical field satisfies the condition $\hbar\omega_{a}/k_{B}T\gg1$ at room temperature. For microwave fields and the nanobeam resonator, whose frequencies are both in the GHz regime, environment temperature in the mK regime --- which can be reached inside a dilution refrigerator --- is sufficient to ensure that $\hbar\omega_{c}/k_{B}T\gg1$ and $\hbar\omega_{b}/k_{B}T\gg1$ \cite{RevMooptomechanicsAspelmeyer,Photonsfri}. To neglect the influences of noise, we assume that our system is operated in the temperature of mK regime, in which the noise terms satisfy the condition $\langle{{{f_{in}}_{+}}}\rangle=\langle{{\xi_{in}}_{+}}\rangle=\langle{{\varsigma_{in}}_{+}}\rangle=0$. Under the mean-field steady-state condition $\langle\dot{\delta{{a_{+}}}}\rangle=\langle\dot{\delta{{c_{+}}}}\rangle=\langle\dot{\delta{{b_{+}}}}\rangle=0$, we obtain
\begin{eqnarray}
\begin{split}
0=(i\lambda_a-\frac{\kappa_a}{2})\langle\delta{a_{+}}\rangle+iG_{om}\langle\delta{b_{+}}\rangle+\varepsilon_{pr},\\
0=(i\lambda_c-\frac{\kappa_c}{2})\langle\delta{c_{+}}\rangle+ig_{em}\langle\delta{b_{+}}\rangle,\\
0=(i\lambda_{b}-\frac{\gamma_{b}}{2})\langle\delta{b_{+}}\rangle+iG_{om}^{\ast}\langle{\delta{a_{+}}}\rangle+ig_{em}\langle\delta{c_{+}}\rangle.
\end{split}
\end{eqnarray}
The solution of $\langle\delta{a_{+}}\rangle$ can be obtained as
\begin{eqnarray}
\langle\delta{a_{+}}\rangle=\frac{\varepsilon_{pr}}{\frac{\kappa_a}{2}-i\lambda_a+\frac{{|{G_{om}}|}^2}{\frac{\gamma_b}{2}-i\lambda_b+\frac{g_{em}^2}{\frac{\kappa_{c}}{2}
-i\lambda_c}}}.
\end{eqnarray}

With the input-output relation of the cavity, the output field at the probe frequency $\omega_{pr}$  can be expressed as \cite{PRAEITAgarwal,PRAPrecisionZhang}
\begin{eqnarray}
\varepsilon_{out}=2\kappa_{a} \langle\delta{a_{+}}\rangle-\varepsilon_{pr}.
\end{eqnarray}
The transmission coefficient $T_{pr}$ of the probe field is also given by \cite{ScienceOptomechanicallyWeis,PRAPhaseJia}
\begin{eqnarray}
T_{pr}=\frac{\varepsilon_{out}}{\varepsilon_{pr}}=2\kappa_{a} \langle\delta{a_{+}}\rangle/{\varepsilon_{pr}}-1.
\end{eqnarray}
Defining $\varepsilon_{T}=\frac{2{\kappa_a}{\langle\delta{a_{+}}\rangle}}{\varepsilon_{pr}}$, we obtain the quadrature $\varepsilon_{T}$ of the output field at the probe frequency $\omega_{pr}$,
\begin{eqnarray}
\varepsilon_{T}=2\kappa_{a} \langle\delta{a_{+}}\rangle/{\varepsilon_{pr}}
=\frac{2\kappa_{a}}{\frac{\kappa_a}{2}-i\lambda_a+\frac{{|{G_{om}}|}^2}{\frac{\gamma_b}{2}-i\lambda_b+\frac{g_{em}^2}{\frac{\kappa_{c}}{2}
-i\lambda_c}}}.
\end{eqnarray}
The real part Re $[\varepsilon_T]$ and imaginary part Im $[\varepsilon_T]$ describe the absorption and dispersion of the system, respectively.

Supposing that both the optical cavity and the microwave cavity are driven at the mechanical red sideband, where $\Delta_{a}^{\prime}=\Delta_c=\omega_{b}$, we define $\lambda=\lambda_a=\lambda_c=\lambda_b$. After the simplification, the term $\varepsilon_{T}$ can be rewritten in a more intuitive form as follows:
\begin{eqnarray}
\varepsilon_{T}=\frac{2\kappa_{a}}{\frac{\kappa_a}{2}-i{\lambda}+\frac{A_{+}}{\lambda_{+}-i{\lambda}}+\frac{A_{-}}{\lambda_{-}-i{\lambda}}},
\end{eqnarray}
where $\lambda_{\pm}$ and $A_{\pm}$ are
\begin{eqnarray}
\begin{split}
\lambda_{\pm}=\frac{\frac{\gamma_{b}}{2}+\frac{\kappa_{c}}{2}{\pm}i{\sqrt{{4{g_{em}}^2-(\frac{\gamma_b}{2}-\frac{\kappa_{c}}{2})^2}}}}{2},\\
A_{\pm}=\pm\frac{\lambda_{\pm}-\frac{\kappa_{c }}{2}}{\lambda_{+}-\lambda_{-}}{{|G_{om}|}}^2.
\end{split}
\end{eqnarray}
This expression has the standard form for the double-OMIT, which is similar to the double-EIT \cite{PRADoubleAlotaibi}.

\section{\label{sec:level1} Physical mechanism of the double-OMIT}
\begin{figure}[tbp]
\begin{center}
\includegraphics[width=0.9\columnwidth]{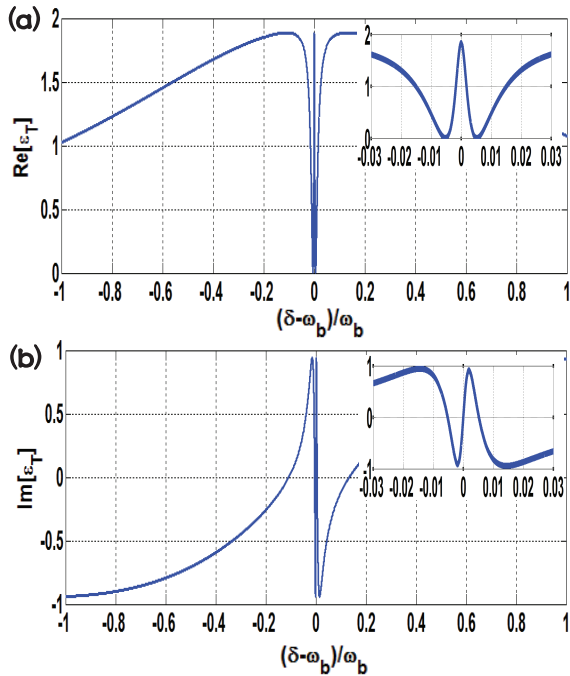}
  \caption{ (color online) (a)The absorption Re$[\varepsilon_T]$ and (b)the dispersion Im$[\varepsilon_T]$ as a function of $(\delta-\omega_{b})/\omega_{b}$. The parameters we chose are $g_{om}/{2\pi}=1.1$ MHz, $g_{em}/{2\pi}=12.3$ MHz, $\kappa_{a}/{2\pi}=5.2$ GHz, $\kappa_{c}/{2\pi}=0.025$ MHz, $\gamma_b/{2\pi}=0.096$ MHz, $\omega_{a}/{2\pi}=194$ THz, $\omega_{c}/{2\pi}=10$ GHz, $\omega_{b}/{2\pi}=2.4$ GHz, $P_{pu}=0.12$ mW, $P_{k}=0.1$ mW, and $\Delta_a=\Delta_c=\omega_b$. The insets in (a) and (b) show the magnified Re$[\varepsilon_T]$ and Im$[\varepsilon_T]$ for the same parametric values, respectively.}
\end{center}
\end{figure}

\begin{figure}[tbp]
\begin{center}
\includegraphics[width=0.9\columnwidth]{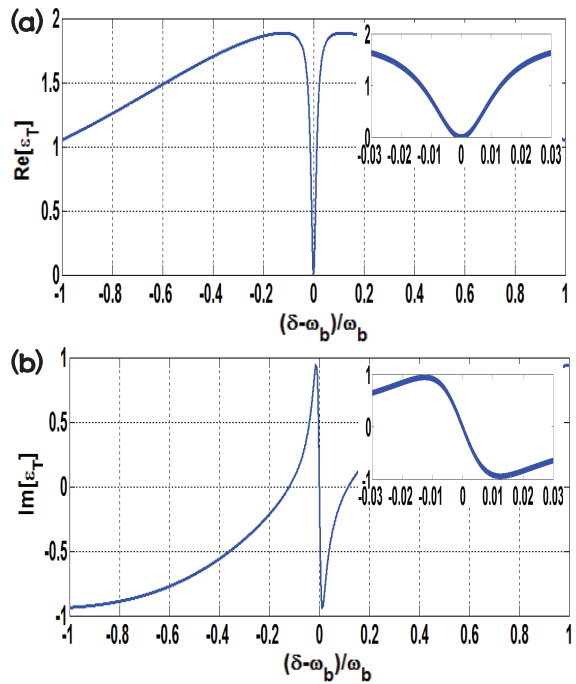}
  \caption{(color online) (a)The absorption Re$[\varepsilon_T]$ and (b)the dispersion Im$[\varepsilon_T]$ as a function of $(\delta-\omega_{b})/\omega_{b}$, with eliminating the piezomechanical interaction, i.e., $g_{em}=0$. The other parameters are the same as those in Fig. 2. The insets in (a) and (b) show the magnified Re$[\varepsilon_T]$ and Im$[\varepsilon_T]$ for the same parametric values, respectively. }
\end{center}
\end{figure}

\begin{figure}[tbp]
\begin{center}
\includegraphics[width=0.9\columnwidth]{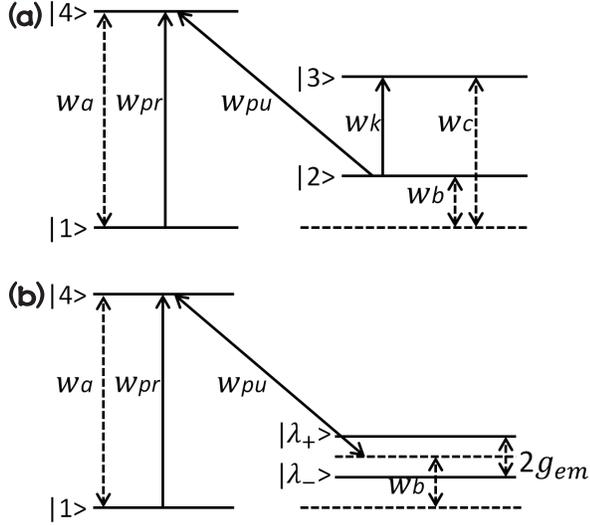}
  \caption{ (a) Energy level structure of the hybrid piezo-optomechanical cavity system, where the states of the levels are designated $|i\rangle (i=1, 2, 3, 4)$. The energy differences between $|{1}\rangle$ and $|{4}\rangle$, $|{1}\rangle$ and $|{3}\rangle$, and $|{1}\rangle$ and $|{2}\rangle$ are the frequencies of cavity $a$, cavity $c$, and mechanical resonator $b$, respectively, and are designated $\omega_a$, $\omega_c$ and $\omega_b$, respectively. $\omega_{pr}$ is equal to $\omega_{a}$, and the detuning events between them are $\omega_{a}-\omega_{pu}=\omega_{c}-\omega_{k}=\omega_{b}$. (b) Energy level structure of the hybrid piezo-optomechanical cavity system in the dressed-state picture. $|\lambda_{\pm}\rangle$ are the new dressed levels generated by the piezomechanical coupling, with the energy difference of $2g_{em}$.}
\end{center}
\end{figure}

We present below a discussion of the feasibility of double-OMIT in the hybrid piezo-optomechanical cavity system. The parameters of the optomechanical crystal AlN-nanobeam resonator we used are based on a realistic system \cite{NatCoherentBalram}, in which the single-photon optomechanical coupling strength can exceed $g_{om}/2\pi = 1.1$ MHz. The parameters of the superconducting coplanar microwave cavity we used are also based on the related experiment, in which the intrinsic quality factor of the superconducting coplanar microwave cavity can exceed $2 \times 10^5$ \cite{APLcavityAM}. The frequency of the  AlN-nanobeam resonator and the decay rate of the optical cavity are $\omega_{b}/{2\pi}=2.4$ GHz and $\kappa_{a}/{2\pi}=5.2$ GHz, respectively, although these values do not strictly meet the condition of the sideband-resolved regime of $\omega_{b}\gg\kappa_{a}$, the assumption that the system is sideband-resolved is still valid \cite{NatCoherentBalram}. The piezoelectric coupling strength between the superconducting coplanar microwave cavity and the optomechanical crystal AlN-nanobeam resonator can be expected to reach $g_{em}/2\pi = 12.3$ MHz for an optomechanical crystal AlN-nanobeam resonator structure with the height, width and length of 0.55 $\mu$m, 1 $\mu$m and 100 $\mu$m, respectively \cite{PRACavityChangLZ}.

We assume that both the optical cavity and the microwave cavity are driven at the mechanical red sideband, where $\Delta_{a}^{\prime}=\Delta_c=\omega_{b}$. The absorption Re $[\varepsilon_T]$ and dispersion Im $[\varepsilon_T]$ of the optical probe field are plotted as function of $(\delta-\omega_{b})/\omega_{b}$, as shown in Fig. 2. In the absorption curve of the optical probe field, two transparency windows can be obtained, for which the positions of two minima points are determined by the imaginary part of $\lambda_{\pm}$, as shown in Eq.(16). The distance between the two minima points is $2g_{em}$, which is closely related to the piezomechanical coupling strength $g_{em}$. If we eliminate the piezomechanical interaction, then the form of $\varepsilon_T$ becomes
\begin{eqnarray}
\varepsilon_{T}=\frac{2\kappa_{a}}{\frac{\kappa_a}{2}-i\lambda_a+\frac{|{G_{om}}|^2}{\frac{\gamma_b}{2}-i\lambda_b}}.
\end{eqnarray}
$\varepsilon_{T}$ has the standard form of the single-OMIT window, as shown in Fig. 3.

This phenomenon originates from the quantum interference effect between different energy level pathways, and the energy level configuration is presented in Fig. 4 (a). In the hybrid piezo-optomechanical cavity system, an N-type four-level system can be formed by the energy levels of the superconducting microwave cavity, the optical cavity and the mechanical resonator. When the input optical fields and the microwave field are applied to the corresponding levels, the energy level of the mechanical resonator is split into two new dressed levels by the piezomechanical coupling effect. Under the condition $\Delta_{a}^{\prime}=\Delta_c=\omega_{b}$ for simplicity, the two new dressed levels are $\lambda_{\pm}$, and the disparity between them is $2g_{em}$, as shown in the dressed-state picture in Fig. 4 (b). Under the effects of the optical radiation pressure, as quantum interference occurs between different energy level pathways, the third-order nonlinear absorption can be enhanced by constructive quantum pathway interference, whereas the linear absorption can be inhibited by destructive quantum pathway interference; as a result, the double-OMIT window appears, and the relevant mechanisms have been studied extensively \cite{PRAObservationYan,PRLCrossPhaseLi}. When eliminating the piezomechanical interaction, no splitting of the mechanical resonator level occurs, and the energy level structure of the scheme remains a $\Lambda$-type three-level system; under the effects of the optical radiation pressure, the double-OMIT window is converted to a single-OMIT window.

\section{\label{sec:level1}Tunable double-OMIT via the optical and microwave fields}
\begin{figure}[tbp]
\begin{center}
\includegraphics[width=0.9\columnwidth]{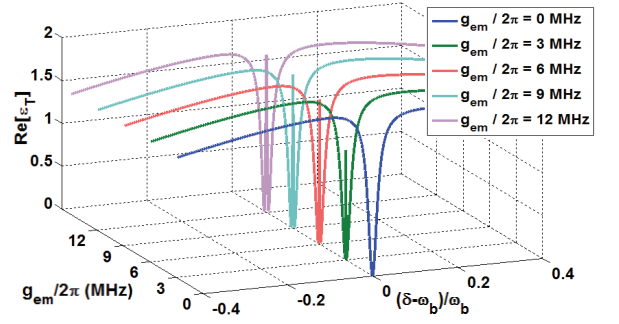}
  \caption{(color online) The absorption Re$[\varepsilon_T]$ as functions of $(\delta-\omega_{b})/\omega_{b}$ and piezomechanical coupling strength $g_{em}$; the units of $g_{em}$ are $2\pi\times 3$ MHz, the values of them are $g_{em}/2{\pi}=0, 3, 6, 9, 12$ MHz, the blue (deep-dark gray) curve, green (dark gray) curve, red (medium gray) curve, cyan (light gray) curve, and purple (shallow-light gray) curve correspond to them, respectively; the other parameters are the same as those in Fig. 2.}
\end{center}
\end{figure}

\begin{figure}[tbp]
\begin{center}
\includegraphics[width=0.9\columnwidth]{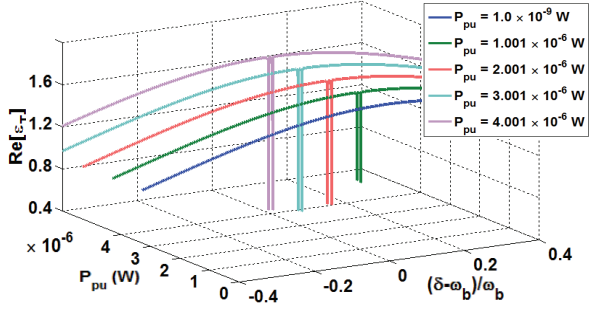}
  \caption{(color online) The absorption Re$[\varepsilon_T]$ as functions of $(\delta-\omega_{b})/\omega_{b}$ and optical pumping power $P_{pu}$; the units of $P_{pu}$ are $1 \times 10^{-6}$ W, the values of them are $P_{pu}=1.0 \times 10^{-9}, 1.001 \times 10^{-6},2.001 \times 10^{-6}, 3.001 \times 10^{-6}, 4.001 \times 10^{-6}$ W, the blue (deep-dark gray) curve, green (dark gray) curve, red (medium gray) curve, cyan (light gray) curve, and purple (shallow-light gray) curve correspond to them, respectively; the other parameters are the same as those in Fig. 2.}
\end{center}
\end{figure}

\begin{figure}[tbp]
\begin{center}
\includegraphics[width=0.9\columnwidth]{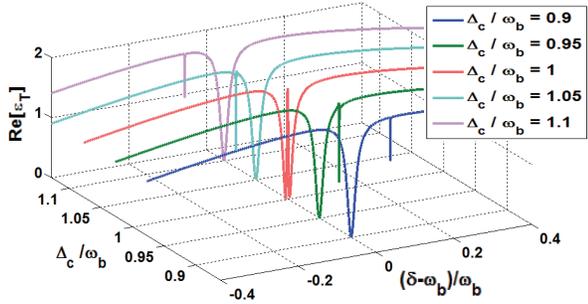}
  \caption{(color online) The absorption Re$[\varepsilon_T]$ as functions of $(\delta-\omega_{b})/\omega_{b}$ and the detuning $\Delta_c$; the units of $\Delta_c$ are $0.05 \times \omega_b$, the values of them are $\Delta_c / \omega_b=0.9, 0.95, 1, 1.05, 1.1$, the blue (deep-dark gray) curve, green (dark gray) curve, red (medium gray) curve, cyan (light gray) curve, and purple (shallow-light gray) curve correspond to them, respectively; the other parameters are the same as those in Fig. 2.}
\end{center}
\end{figure}

\begin{figure}[tbp]
\begin{center}
\includegraphics[width=0.9\columnwidth]{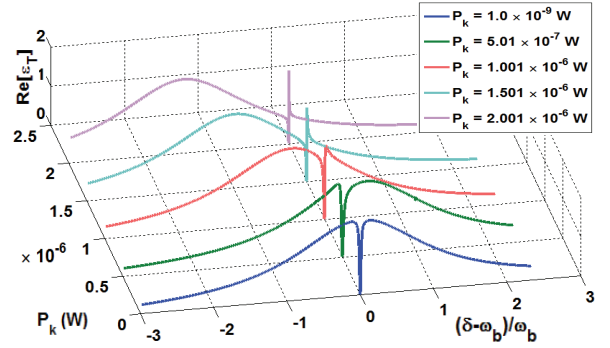}
  \caption{(color online) The absorption Re$[\varepsilon_T]$ as functions of $(\delta-\omega_{b})/\omega_{b}$ and microwave field strength $P_{k}$, with $\Delta_c=0$; the units of $P_{k}$ are $5 \times 10^{-7}$ W, the values of them are $P_{k}=1.0 \times 10^{-9}, 5.01 \times 10^{-7},1.001 \times 10^{-6}, 1.501 \times 10^{-6}, 2.001 \times 10^{-6}$ W, the blue (deep-dark gray) curve, green (dark gray) curve, red (medium gray) curve, cyan (light gray) curve, and purple (shallow-light gray) curve correspond to them, respectively; the other parameters are the same as those in Fig. 2.}
\end{center}
\end{figure}

To further explore the characteristic of the microwave-controlled optical double-OMIT, we plot the absorption Re$[\varepsilon_T]$ as functions of $(\delta-\omega_{b})/\omega_b$ and $g_{em}$. As shown in Fig. 5, in the absence of the controlled microwave field, only a single transparency window of the optical probe field appears at the frequency $\delta=\omega_{b}$. With the enhancement of the piezomechanical coupling strength, the single transmission window is split into two transparency windows, and the separation between them increases. This phenomenon originates from the splitting of the mechanical resonator that is determined by the piezomechanical coupling strength, in which the positions of the two minima points corresponds to the imaginary part of $\lambda_{\pm}$. When the piezomechanical coupling strength is zero, there is no splitting of the mechanical resonator, and the energy level structure of the scheme remains a $\Lambda$-type three-level system; under the effects of the optical radiation pressure, the single-OMIT is obtained. With the enhancement of the piezomechanical coupling strength, the mechanical resonator is split into two new dressed levels $\lambda_{\pm}$, and the $\Lambda$-type three-level structure of the scheme is replaced by the N-type four-level structure; with the occurrence of the double-OMIT, the single transmission window is also split into two transmission windows.

In the situation shown in Fig. 5, as the transmission rate of the probe field at the frequency of $\delta=\omega_{b}$ is determined by the piezomechanical coupling strength, this phenomenon can be used to realize an optical switch via the modulation of the piezomechanical coupling strength from zero to a non-zero value, based on the relation that the strength is inversely proportional to the distance between the top electrode and the AlN-nanobeam resonator; and this phenomenon is similar to the optical switch realized in natural atomic systems \cite{PRAswitchingKumar}. When the two transparency windows appear, a narrow absorption line can be obtained at frequency $\delta=\omega_{b}$. As the linewidth of the narrow absorption line is approximately equal to $g_{em}$, which is small enough relative to the linewidth of the optical cavity ($\kappa_{a}/{2\pi}=5.2$ GHz), this phenomenon can be potentially applied in the field of high-resolution spectroscopy and is similar to the sub-Doppler spectral resolution observed in natural atomic systems \cite{PRLNonlinearHarris}. Furthermore, when the double-OMIT occurs, our system is in a dark state \cite{PRAdarkstate}, in which the population of the system is in a coherent superposition of the states $|{1}\rangle$, $|\lambda_{+}\rangle$ and $|\lambda_{-}\rangle$. This is similar to the single-OMIT, in which the population is in the coherent superposition of the states $|{1}\rangle$ and $|{2}\rangle$. As a result, our system also provides an approach for coherent population trapping in solid-state quantum systems.

Fig. 6 presents the absorption $Re[\varepsilon_{T}]$ with respect to $(\delta-\omega_{b})/\omega_{b}$ for different strengths of the optical pump field. We find that, in the absence of the optical pump field, when only an intense microwave field is applied to the system, no transparency windows appear, because without the optical radiation pressure, even though the microwave field is intense enough to generate strong piezomechanical coupling, no quantum interference occurs between different energy level pathways. With increasing optical pump field power, the transmission depth of the double-OMIT window also increases. This phenomenon occurs because the transmission rates of the two minima points of $A_{\pm}$ are both proportional to the optomechanical coupling strength $|G_{om}|$, as shown in Eq.(15); with the enhancement of the optomechanical coupling strength, the quantum interference strength between different energy level pathways becomes increasing strong, thereby increasing the transmission depth of the double-OMIT window. In this situation, as shown in Fig. 6, the transmission rates of the probe field at two minima points can both be adjusted by the optical pump field, and the frequencies of the probe field at those two minima points can also be used as the information channel, hence this phenomenon can potentially be applied in double-frequency channel information processing in solid-state quantum systems.

Furthermore, we discuss the situation in which $\Delta_c$ is different from the mechanical resonator frequency $\omega_b$, where $\Delta_c$ is the detuning of the microwave driven field from the microwave cavity. As illustrated in Fig. 7, when the detuning $\Delta_c$ is far from the frequency of the mechanical resonator $\omega_b$, no double-OMIT appears. When the detuning $\Delta_c$ is slightly different from that of $\omega_b$, the asymmetric double transparency window appears. Relative to the case of $\Delta_c=\omega_b$, the absorption curves move rightward (leftward) in the case of $\Delta_c<\omega_b$ ($\Delta_c>\omega_b$). With an increasing difference between $\Delta_c$ and $\omega_b$, the degree of window asymmetry also increases. This phenomenon is characterized by Eq.s (14)-(16), when $\Delta_c \neq \omega_b$, the condition $\lambda_a=\lambda_c=\lambda_b$ converts into $\lambda_a \neq \lambda_c \neq \lambda_b$. As a result, the mechanical resonator level splits into two new asymmetric dressed levels, with the separations between them from the mechanical resonator level and the transmission rates of the two minima points being unequal. In this situation, because the frequencies of the two minima points can both be controlled by changing the frequency detuning between the microwave driven field, and because the frequencies of the probe field at those two minima points can be used as the information channels, this phenomenon can also be applied for tunable double-frequency channel information processing in solid-state quantum systems.

For a more comprehensive treatment, we consider the condition that the frequency of the microwave field is resonant with the superconducting coplanar microwave cavity, for which the detuning $\Delta_c=0$. As shown in Fig. 8, when a weak controlled microwave field is applied to the system, only a single transparency window appears at the frequency of $\delta=\omega_b$. With the enhancement of the controlled microwave field power, the single transmission window gradually becomes a narrow-line. This phenomenon arises because when $\Delta_c$ is far from $\omega_b$, based on Eq.s (14)-(16), the mechanical resonator level splits into two new asymmetric dressed levels, for which both the separations between them from the mechanical resonator level and the transmission rates of the two minima points are unequal. When the detuning $\Delta_c=0$ is too far from $\omega_b$, the asymmetry degree of the window is so high that the smaller of the two transparency windows vanishes. With the enhancement of the microwave field, based on Eq.(5), the effective detuning between the optical pump field and the optical cavity becomes increasingly greater than $\omega_{b}$; as a result, the optomechanical coupling strength $|G_{om}|$ decreases gradually, causing the single transmission window to shrink into a narrow-line. In this situation, because the transmission rates of probe field at frequencies $\delta=\omega_b$ can be controlled by changing the power of the microwave driven field, similar to the situation shown in Fig. 7, this phenomenon can also be used for tunable single-frequency channel information processing in solid-state quantum systems.

\section{\label{sec:level1} Conclusion}
In conclusion, our proposed scheme provides a feasible way to control the optical field with a microwave field in solid-state quantum systems, in which a piezoelectric optomechanical crystal resonator is placed in a superconducting microwave cavity. In this scheme, an N-type four-level structure can be formed. Under the effects of the radiation pressure and the piezoelectric interaction, the quantum interference between different energy level pathways induces the occurrence of the double-OMIT. Similar to the double-EIT observed in natural atomic systems, the double-OMIT can also be applied in the fields of optical switches, high-resolution spectroscopy, coherent population trapping and quantum information processing. With the advantage of integration, our system can be extended to other hybrid solid-state systems, which would be helpful for exploring new quantum phenomena.

\begin{acknowledgments}
This work is supported by the Strategic Priority Research Program (Grant No. XDB01010200), the Hundred Talents Program of the Chinese Academy of Sciences (Grant No. Y321311401), the National Natural Sciences Foundation of China (Grant Nos. 11347147, 61605225, 11674337 and 11547035), and the Natural Science Foundation of Shanghai (Grant No. 16ZR1448400).
\end{acknowledgments}

\bibliographystyle{apsrev4-1}
\bibliography{doubleOMIT}

\end{document}